\documentstyle[preprint,aps,psfig,rmp]{revtex}
\tightenlines
\begin{document}
\title{Why the Universe is Just So}
\author{Craig J. Hogan}
\address{Astronomy and Physics Departments \\
University of Washington\\
Seattle, Washington 98195, USA}

\maketitle

\begin{abstract}
 Some properties of the world are fixed  by physics derived from
mathematical symmetries,  while others are selected from an
ensemble of possibilities. Several     successes and failures of 
``anthropic'' reasoning in this context are reviewed 
in the light of recent developments in astrobiology, 
cosmology and   unification physics.
Specific issues raised include our spacetime location
(including the  reason for the present age
of the universe), the timescale of biological evolution,  
the tuning of global  cosmological parameters, and
the origin of the Large Numbers of astrophysics  and the 
parameters of the Standard Model.
Out of the twenty parameters of the Standard Model,
the basic behavior and structures of the world (nucleons, nuclei,
atoms,   molecules, planets, stars, galaxies) depend  mainly on  five 
of them: $m_e,m_u,m_d,\alpha,$ and $\alpha_G$ (where $m_{proton}$ and
$\alpha_{QCD}$ are taken as defined quantities). Three of these
appear to be independent in the context of Grand Unified Theories
(that is, not fixed by any known symmetry) and at the same time have values
within a very narrow window which provides for stable nucleons and
nuclei and abundant carbon.  The conjecture
is made that the  two light quark masses
 and one coupling constant  are ultimately determined
even in the ``Final Theory'' by  a choice from a large
or continuous ensemble, and the prediction is offered  that the
correct unification scheme will not allow calculation of
$(m_d-m_u)/m_{proton}$ from first principles alone.
\end{abstract}
\break

{\it ``What really interests me is whether
God had any choice in creating the world.''}\footnote{Einstein
is also famous for declaring   that ``God
does not play dice''.  This comment did not refer to the structure of 
physical law
but to the randomness and indeterminacy inherent in quantum measurement.}

---Einstein
\bigskip
\bigskip
\tableofcontents

\section{Necessity, Chance and Selection in Physics}

Which things about the world are accidental, which things
are necessary? Philosophers have debated this metaphysical question  for
thousands of years (see, e.g., Leslie 1989), but  it
has become more than an abstract  philosophical issue since the answer  now  
influences  the mathematical  design of fundamental physical theory
(see, e.g., Tegmark 1998).
Within the confines of physics
we can sharpen the question
and can even hope to offer some provisional answers.

The question now has special  currency because in modern
fundamental theories,   low-energy
effective  constants can 
preserve the symmetry of    precise
 spatial uniformity  over a large 
spatial volume--- even a whole ``daughter universe''---
even while they adopt different
values in different universes.
In addition,  inflationary cosmology offers a physical
mechanism for creating a
true
 statistical
ensemble (a ``multiverse''; Rees 1997)
where many possible values of the constants are
realized.  The truly fundamental equations may be
the
 same everywhere in all universes
but may not completely determine the values of all the effective, 
apparently  ``fundamental'' constants
at low energies in each one. The
 Theory of Everything currently under construction, 
even in its final form,
 may never provide a derivation
from  first principles of all the pure numbers controlling everyday
phenomenology. These may instead be primarily determined  by a kind of
selection, dubbed  the ``anthropic principle'' by Carter,
the ``principle of complexity'' by Reeves, the ``principle of 
effectiveness'' by Rozenthal, such that the elementary
building blocks  of
the universe allow  for complex things to happen,
such as  the assembly  of observers.
We can seek clues to
the flexible degrees of freedom in the ``final theory''
by looking for parameters of the effective
low-energy theory  (the Standard Model) 
with especially powerful effects: parameters whose small
variation from their actual fortuitous values lead to 
major qualitative changes.

Since the   reviews of 
Carr and Rees (1979) and Barrow and Tipler (1986), advances in both physics and
astronomy have,  amazingly, led to progress on the ancient riddle of
chance and necessity, on very different fronts:
at one extreme   the very
concrete circumstances about our local habitable environment
and its detailed history; at the
other extreme,  the most abstract levels of physics.
 The natural history of the
solar system and the Galaxy have revealed new
couplings between biology and the astrophysical environment,
as well as actual data on other solar systems.
 Inflationary multiverses    (e.g.,   Vilenkin   1998b)
 now provide a physical
framework to discuss  
  different   choices of physical vacuum which may allow   some
of the parameters of low-energy physics (which we try to identify)
to be tuned by selection.
At the same time,   unified theories
  constrain some  relations among
the parameters to be fixed by symmetry.
{\it  Remarkably, the freedom still available to tune
 parameters in Grand Unified Theories  
 appears well matched to that required to select parameters
which yield  a
complex phenomenology at low energy.}
 Simple arguments suggest that one independent 
coupling constant and   two out of the three light fermion 
 masses (the   down quark mass, and either the up quark or
 electron mass)  may  not be fixed by symmetry, which allows the 
fundamental theory enough flexibility to find a combination with a
rich nuclear and
chemical phenomenology;
 the  other relationships among the 20 or more 
 parameters of current standard theory
can be fixed by symmetries of unification mathematics.

It is easy to guess wrong about selection effects and it is worth recalling
the history   of the Large Numbers Hypothesis.
Dirac (1937)
saw two of 
the large numbers of nature--- the weakness of gravity and
the low density  of the universe--- and concluded, incorrectly, that
gravitational coupling depends on cosmic density. The correct insight
(by Dicke, 1961) was that the density of the universe is determined
by its age, and the age of the universe is mainly fixed
by our own requirements, probably mainly to do with how
long it takes stellar populations to synthesize the heavy nuclei
needed for planets and life. The long timescales associated
with stars ultimately derive from the weakness of gravity and the 
energy available from nuclear fusion.
Once it is granted that our presence requires evolved stars,
Dirac's coincidence can be derived from physical models of stars. 
Carter (1983) extended the argument to draw conclusions about the
intrinsic
timescales of biological evolution, some of which appear 
to be confirmed by modern astrobiology. Fossil evidence now confirms
 intricate couplings of biological and astronomical
processes throughout the history of the Earth, and we have developed 
enough understanding to guess that highly complex life requires
a  rare combination of  factors (Ward and Brownlee 1999).

It is also easy to discredit anthropic arguments. In the same
way that Darwinian natural selection can be discredited by
silly ``Just So Stories'' (How the Leopard Got His Spots, etc.),
anthropic arguments are sometimes used indiscriminately;
for example, when a theory of quantum cosmology essentially fails to predict
anything, so that all the important features of the universe must
be attributed to selection. Such extreme applications of
anthropic reasoning  undermine the essential goal
of unification physics, to achieve an elegant mathematical explanation
for everything. Yet one must bear in mind--- dare we call it
a Principle of Humility?--- that at least some properties of
the world might not have an elegant mathematical explanation,
and we can try to guess which ones these are.

\section{Our Location in Spacetime}

\subsection{Why the Universe is   Old}

The large-scale character of spacetime is well established to be a
large, nearly homogeneous, expanding  3-space with 
a (real or imaginary) radius of curvature vastly larger
 than any microscopic scale.
This fundamental structure, which used to seem to require fine
tuning of initial conditions, is now understood as a natural causal
consequence of inflation, which  automatically creates macroscopic spacetimes,
exponentially larger than microscopic  scales, from 
microscopic instabilities.

Our time coordinate in this spacetime,
  now estimated to be about 12 to 14 Gy, is 
(as Dicke argued) probably selected by our own needs. The simplest
of these is the need for
a wide variety of chemical elements.
The early universe produced nearly pure
hydrogen and helium, but
biochemistry uses almost all of the  chemically active,
reasonably abundant elements in  the upper half of the periodic table.
The time required to manufacture abundant
 biological elements and stars with earthlike
planets is determined by the formation and evolution  times of galaxies and
stellar populations,
setting a minimum age of 
   billions  of years.

Curiously,  most  observations now suggest that  we also appear
 to be living at an intrinsically 
special time in the history of the expansion.
Data on the Hubble constant, the age of the universe, 
cosmic structure, matter density, and in particular the   supernova Hubble
diagrams of  Riess et al. (1998) and Perlmutter et al. (1999),
and   microwave background anisotropy, e.g. Miller et al. (1999),
de Bernardis et al. (2000), Hanany et al. (2000),
all support a cosmological model with close to a spatially flat geometry, a low
matter density, and a significant component of ``dark energy'' such as
a cosmological constant (see Fukugita  2000 for a review of the data).
These models  have  an intrinsic expansion
rate  $(\Lambda/3)^{-1/2}$ introduced by the cosmological
constant $\Lambda$, which happens   to be comparable
to the current Hubble rate $H_0$. The rough coincidence of this
fundamental scale,  fixed by the  energy density of
the physical vacuum  $\rho=\Lambda/8\pi G$, with   seemingly unrelated 
astrophysical timescales determined by stellar evolution,  has invited
anthropic explanations (Weinberg 1987,1989, 1997, Vilenkin 1995,
Efstathiou 1996, Martel et al 1998, Garriga et al. 2000). 

The conjecture is that
in a large  ensemble of universes (a multiverse), most universes have very large
values of the cosmological constant which render them uninhabitable; 
the value we observe is not the most probable one but is typical of that seen by the
largest number of  observers in the multiverse as a whole. 
This argument
is tied up with another parameter, the amplitude of the fluctuations
which produce galaxies, now usually thought to be determined
by the detailed shape of the potential controlling cosmological
inflation (e.g. Kolb 1996), which may also be determined by 
selection (Tegmark and Rees 1998).
The anthropic prediction of cosmological parameters in multiverses is   still tied up 
in the murky unresolved debates of quantum cosmology which describe the 
ensemble
(Turok and Hawking 1998, Vilenkin 1998a, Linde 1998).

The value of $\Lambda$ need not be set anthropically.
A similar   exotic form of dark
energy (``Quintessence''), a dynamical scalar field
with properties controlled  by an
internal potential, could  evolve   in such a way as to adjust  to  give it a density
comparable to the matter density today (e.g. Zlatev et al. 1999).  
Or perhaps  a ``derivable''
fundamental scale of physics exists,    
 corresponding to a vacuum  energy
density which happens to be 
  about the same as the current cosmic mean density.
The current cosmic mean density $(\simeq 0.1{\rm mm})^{-4}
\simeq (0.003 {\rm eV})^{4}$
is derivable from Dicke's argument in terms of fundamental constants;
the required coincidence  (see 
equation \ref{eqn:salpeter} below) is that
 $\Lambda\approx \left({m_{Planck}/
m_{proton}}\right)^6 t_{Planck}^2 $.   

One way or another, the intrinsic global cosmological parameters are
intimately connected with the large numbers
(or ``hierarchy problem'')  of fundamental physics; but
the nature of the connection is still not clear. 

\subsection{Why the Universe is Just So Old}

Why is the universe not much older than it is? In the
anthropic view, part  of the reason
must be the decrease in new star formation, which both globally 
and within the Galaxy has
decreased by almost an order of magnitude during the 4.5
billion years since the solar system formed (Fukugita et
al. 1996, Lilly 1998, Madau 1999). Galaxies
have converted the bulk of their original gas to stars or 
 ejected it altogether, and
the larger reservoir of intergalactic gas is now too hot
to cool and collapse to replenish it (Fukugita et al 1998,
 Cen and Ostriker 1999).

The decrease in star formation rate
also means  that the heavy element production rate is
decreasing,
 and therefore the mean age of radioactive elements
(especially those produced by Type II supernovae, whose
rate is closely tied to current star formation) is 
increasing. The   new planets which  are  forming now
and in the future are  less radioactively alive  than Earth
was when it formed.
Since  abundant
live radioactive nuclei in the Earth's core 
(especially uranium 238, thorium 232, and  potassium 40,
with half-lives of 4.46, 14 and 1.28 Gy respectively) are needed to
power vulcanism, continental drift, seafloor spreading,
 mountain uplift, and the convective dynamo which creates the 
Earth's magnetic field, new
planets even in the rare instances where
they do manage to form will in the future  not have these
important attributes of the Earth.  Life is also sensitive
to other features of  the detailed composition inside the Earth: the
 correct iron abundance
is needed to provide sufficiently    conductive core flows to
give a strong magnetic field. Without its protection, the solar wind
would erode the atmosphere as it appears to have done on Mars
since the magnetic dynamo ceased there (Acu\~na et al. 1999, 
Connerney et al. 1999).
The coupling of bioevolution with astrophysics thus defines a fairly sharp
window of habitability in cosmic time 
as well as space (Ward and Brownlee 1999):
New stars and  new habitable planets   are becoming increasingly 
rare.

\subsection{Coupling of Biological and Astronomical Timescales}

We find more specific clues to factors influencing our time
coordinate by a closer examination
of local natural history, both in the fossil record (e.g. Knoll 1999) and the 
genomic one (e.g. Woese et al. 1990, Doolittle 1999). The
oldest sedimentary
 rocks (from 3.9 Gya, where Gya=$10^9$ years ago) on the surface of the Earth are almost
as old as the Earth itself (4.55 Gya), yet appear to
harbor
 fossilized cells. Unambiguous fossils of 
cyanobacteria, closely resembling modern species, are found from 3.5 Gya.  The earliest  
life seems to have emerged  soon (within of the order of 0.1 Gy) after the  last 
globally-sterilizing meteoroid impact.  The
first   eukaryotic fossils ({\it Grypania}) 
show up much  later, about 2Gya (about the time  when the atmospheric oxygen 
level rose substantially); widespread eukarya (acritarchs, a form of planktonic algae)
do not appear 
until much more recently (1.5 Gya).  Significant
morphological diversity only began about 1 Gya, possibly paced by the emergence of
sex. The   Cambrian explosion, which took place over a remarkably narrow interval 
of time  between about 0.50 and  0.55 Gya,  created essentially all of the
variety and complexity in body plans of modern animals.   Since
then there have been  several  
 mass extinctions triggered by catastrophic impacts (including possibly the huge
Permo-Triassic event   0.25 Gya, and almost certainly the smaller dinosaur killer
Cretaceous-Tertiary (KT)
 event   0.065
Gya), indicating that extraterrestrial factors are even recently  at work
in shaping biological history.

What is the clock that determines the roughly 4 Gy timescale
from the formation of the Earth to the Cambrian explosion? 
If it is a purely biological clock, there is a striking coincidence between
this timescale and the main-sequence lifetime of the Sun,
about 10Gy. 
Why should Darwinian bioevolution occur on a similar timescale
to stellar evolution?
Why should it be that we show up when the Sun is just
halfway through its lifetime?
Carter (1983) considered these coincidences and proposed an
anthropic explanation: if   the biological
clock has a very long intrinsic timescale, most systems fail
to evolve significantly before their suns die; those that by chance evolve
quickly enough will tend to do so ``at the last minute''. If there
are a small number of rare rate-limiting steps, the coincidence
can be explained.

Indeed  the emerging picture of continual cosmic catastrophes
affecting the biosphere 
and the mounting evidence for the intimate coupling of life and
the global environment has started to flesh out the details of
what paced evolution, and how it has been 
controlled or limited by astrophysical events and thereby by astrophysical timescales.
In addition to asteroid and comet impacts,
intimate   couplings are now recognized
between geophysical and biological evolution, although
their relative importance is not settled.

 One 
example is the   
global carbon cycle,
which includes biological components
(important in the precipitation of carbonates)  
as well as plate tectonics, vulcanism, and climate-controlled 
erosion; the sum of these elements 
may allow  the planet
to maintain a surface temperature which tracks the habitable zone,
in spite of variations in insolation since the Sun formed
of up to twenty percent (Schwartzmann and Volk 1989).
The most spectacular failures of this stabilization mechanism may have 
led to   ``Snowball Earth" events (Evans
 et al. 1997, Hoffman  et al. 1998) where the entire surface of the planet
iced over, and the subsequent superheated recovery from these events by volcanic
replenishment of  greenhouse gases.  The most recent of these events may have 
triggered the Cambrian explosion. 
Another example is the accumulation of oxygen, a biological process
partly paced by geochemistry (the global  oxidation of iron) which also
took place over a billion years, which  certainly
 enabled  and may have paced the
 explosion of complex life forms.

 Direct evidence thus suggests that 
interdependent ``co-evolution'' accounts for the coincidence
of biological and astrophysical timescales, even though the dominant
couplings may not yet be known. The actual situation is subtly different from
Carter's original guess; the intrinsic timescale of biological evolution, if
one exists, appears to be relatively  rapid, and the pace of evolution
 has been set by occasional rare opportunities (such as the isolation of
Darwin's finches on various Galapagos islands, but on a global scale).   Carter's 
main conclusion, that advanced life  is  relatively rare, is substantiated by
the accumulation of evidence over the last twenty years:  
many fortuitous circumstances seem to have played a role
in the emergence of animal life on Earth (Ward and Brownlee 1999).\footnote{As
yet another example, Gonzalez (1999)
has recently pointed out that even the orbit of the solar
system in the Galaxy appears to  be finely tuned to reduce comet
impacts:
compared to other stars of the same age,
the sun  steers an unusually quiet path through the Galaxy---
an orbit with unusally low eccentricity   and small amplitude of
vertical motion out of the disk. This could be explained
anthropically,
perhaps through the effect of 
 Galactic tidal distortions on the Oort comet
cloud which create catastrophic storms of comet impacts in the inner
solar system (Heisler and Tremaine 1986, Matese and Whitmire 1996).}

\section{Fixed and Tunable Parameters of   Physics}

\subsection{The Standard Model and Everyday Life}

The  Standard Model of fundamental quantum fields has at least
20 adjustable parameters (including for this count
Einstein's classical theory of gravity), although it 
explains almost all natural phenomena with less than half of these,
and the basic structures are fixed by just a handful of them.
At a deeper level, the values of the  parameters 
are presumed to be not all   truly independent
 and adjustable;   symmetries
fix relationships between some of them.

The minimal Standard Model has 19 ``adjustable'' parameters
(Cahn 1996, Gaillard et al. 1999): Yukawa coefficients
fixing the  masses of the six quark
and three lepton flavors ($u,d,c,s,t,b,e,\mu,\tau$), the Higgs mass
and vacuum expectation value $v$ (which multiplies the
Yukawa coefficients to determine the fermion masses), three angles and one
phase of the CKM (Cabibbo-Kobayashi-Maskawa) matrix (which mixes quark weak- and
strong-interaction eigenstates), a phase for the quantum chromodynamic (QCD) vacuum,
and three coupling constants $g_1,g_2,g_3$ of the  
gauge group, $U(1)\times SU(2)\times SU(3)$.
 If as seems likely the neutrinos are not massless, there
 are seven more parameters for them
(three masses and another four CKM matrix elements). 

Various more or less observable combinations of these parameters 
appear in discussing phenomenology, taking account of
the change of couplings with energy. 
The traditional zero-energy electromagnetic fine structure constant
$\alpha=e^2=1/137.03599$,
changes with energy scale to $\alpha(m_Z)\approx 1/128$
at the $Z$ mass scale; it is related to the electroweak parameters
by
$e=g_2\sin\theta_W$, where 
the weak mixing angle $\tan \theta_W\equiv g'/g_2$
also fixes the $W$ and $Z$ mass ratio,
$sin^2 \theta_W=1-(m_W^2/m_Z^2)$, and for consistently 
normalized currents one defines
$g_1=\sqrt{5/3}g'$. 
The Fermi constant of weak 
interactions can be written 
\begin{equation}
G_F={\sqrt{2} g_2^2\over 8 m_W^2}
={\sqrt{2}\over 8}{\alpha\over m_W^2\sin^2\theta_W}
={v^2\over\sqrt{2}}
\end{equation}
where $v=246$GeV is the expectation value of the Higgs
field.
The strong
coupling $\alpha_S\equiv g_3^2$ can be defined 
 at some energy scale $\Lambda$, say
$\alpha_s(\Lambda=m_Z)=0.12$; or, an energy scale $\Lambda_{QCD}\approx 200$MeV
can be defined where the coupling diverges. The masses of protons
and other hadrons are  thereby approximately  fixed by the value of $\alpha_S$ at any
energy.\footnote{The relation of $\Lambda_{QCD}$ to $\alpha_S(E)$
 also depends on the fermion and Higgs
masses, through threshold effects.}

 The Standard
Model plus classical gravity describes all known physical
phenomenology
 within the
present-day universe.  Everyday matter  (indeed nearly
all
 of the ``baryonic'' matter of the universe aside from energetic particles)
 is almost entirely made of the lightest   first generation
fermions.\footnote{The higher generations are less prominent in nature than 
the first because they are 
heavier and decay by weak interactions, although  they are always
present at some level because of mixing and probably play important roles in
supernova physics and other exotic but important astrophysical environments
such as neutron stars. They also enter through the CKM matrix,
one  complex phase of which  is a source of observed CP violation
and therefore possibly 
related to the physics responsible for creating the cosmic
excess of matter
 over antimatter. The masses of the heavy
fermions matter little to
familiar natural phenomenology, so they  could 
 be set by the choices selectively adopted by the
 first generation if the fermion masses of the three generations
 are (as is conjectured) coupled
to each other in a unified scheme  by a
 mixing matrix. 
There are many such schemes proposed (Berezhiani 1996); for example, 
in the ``democratic'' scenario of 
 Fukugita, Tanimoto and Yanagida (1999), the nine fermion masses
are determined by five parameters, and still only two independent parameters
determine the masses of $u,d,e$ (with $m_d/m_e$ fixed by SO(10)).}   
Since we may take the strong coupling to be fixed at
the proton mass scale, and the fermion masses enter mostly through their
ratio to the nucleon mass, the basic structures (almost) just depend  on
four   parameters, which we may take to be
the three  light fermion masses $m_e,m_u,m_d$ and the elecromagnetic
coupling constant
$\alpha$, plus gravity.\footnote{Agrawal et al. (1998) have developed the  point of view
that the weak scale itself is determined anthropically  and that $v$ is the one  tunable
parameter--- singled out in the standard model  by having a dimension.
 Indeed the  fundamental  degrees of freedom of the
fundamental
theory are not known and one of the main objectives of studies such
as these is to sniff them out. Here I imagine adjusting some 
coefficients  in the Lagrangian according to the constraints imposed
by unification. This amounts to exploring a different space of
variation, with more degrees of freedom, than Agrawal et al.
For most of the arguments presented here, it does not matter whether
the Higgs is counted as a separate degree of freedom.
Note however that tuning only the Higgs varies all the fermion masses
in lockstep, and cannot by itself tune more than one degree  of freedom.} 
Newton's constant of universal gravitation $G$ 
specifies the 
coupling of all forms of energy to gravity (which is usually regarded as
outside the ``Standard Model'').
In the next section we review how the gravitational coupling of nucleons
$Gm_{proton}^2$   defines the  relationship between the structure of the
astronomical scales of the universe and those of the microworld.

  The  electron
mass and fine structure constant together determine
  the  basic behavior of atomic matter and its interaction
with radiation--- in other words, all of chemistry and biology.
They enter  in familiar combinations
such as the  classical electron radius   $r_e=\alpha/m_e$,
the Thomson cross section   $\sigma_T=(8\pi/3)(\alpha/m_e)^2$,
the electron Compton wavelength $\lambdabar_e=m_e^{-1}$, and
the Bohr radius   $a_{Bohr}= (\alpha m_e)^{-1}$.
The Rydberg energy $m_e\alpha^2/2$ sets the scale of atomic binding; atomic
fine structure (spin-orbit) 
splittings depends on  higher powers of $\alpha$, and splittings of
 molecular modes, which
include electronic,
vibrational and rotational states, depend on powers of
$m_e/m_{proton}$.

The detailed  relationships among atomic and molecular  eigenstates are not preserved
continuously or homologously as $\alpha$ and $m_e$ are adjusted,
and   would be scrambled with
even small changes. 
However,  structural  chemistry would not change much 
if $\alpha$ and $m_e$ were adjusted slightly  differently. 
The structure of electron orbitals in atoms and molecules scales
homologously in first order with the Bohr radius, and the energy
levels of the ground-state orbitals scale with the Rydberg. So,
while it does seem miraculous that complementary structures
can form with the  specificity (say) of purines and pyramidines in DNA,
the possibility of this
 miracle can be traced back to  group theory and quantum mechnanics;
if $\alpha$ and/or  $m_e$ changed, the DNA structure would remain
almost the same,
it  would just change size relative to, say, the classical electron radius. 
(The departure from homology enters only in subdominant
terms in the Hamiltonian, such as the spin-orbit or
 nucleus-nucleus interactions.)

This amazing achievement of quantum theory illuminates  another  good example
of failed anthropic reasoning. Before quantum mechanics, it was suggested  that atomic
properties must have been  tuned 
to achieve the marvellous  chemical structures
needed for life (Henderson 1913). Instead
it appears that ordinary Darwinian natural selection has 
found and exploited the structural   opportunities
presented by underlying symmetries. Biology and 
not physics or cosmology  should be given credit for this miracle!

By contrast, 
changing the quark masses even a small amount  has   drastic 
consequences which no amount of Darwinian selection can
compensate.
The $u-d$ mass difference in particular attracts attention  because the
$d$   is just enough heavier than  $u$ to overcome the
electromagnetic energy difference to make the proton ($uud$) lighter
than
 the neutron
($udd$) and therefore stable. On the other hand if it were a little
heavier still, the deuteron would be unstable and it would
be difficult to assemble any  nuclei heavier than hydrogen.
This then is a good candidate for selective tuning among
multiverses. Similarly, 
the sum of the quark  masses controls the pion mass,
so changing them alters the
 range of the nuclear potential and significantly changes nuclear structure
and energy levels. Even a small change radically alters the 
history of nuclear astrophysics, for example,
by eliminating critical resonances of nucleosynthesis 
needed to produce abundant carbon (Hoyle 1953).
 It would be surprising if   symmetries
 conspired to
satisfy  these constraints, but quite natural if
the parameters can adopt a continuous range of values.
{\it One therefore expects these  particular parameters to
continue to elude relationships
fixed by symmetries.}

\subsection{Structures and Timescales of the Macroworld}

Essentially  all astrophysical structures,
sizes and timescales are controlled by one 
dimensionless ratio,
sometimes called the ``gravitational coupling constant,''
\begin{equation}
  \alpha_G  ={ {G m_{proton}^2}\over \hbar c} =
\left({m_{proton}\over m_{Planck}}\right)^2
\approx 0.6\times 10^{-38}
\end{equation}
where $m_{Planck}=\sqrt{\hbar c/G}\approx 1.22\times 10^{19}$ GeV
 is the Planck mass 
and $G=m_{Planck}^{-2}$ is Newton's gravitational constant.\footnote{The 
Planck time $t_{Planck}=\hbar/m_{Planck}c^2=m_{Planck}^{-1}=0.54\times 
10^{-43}$sec is the
quantum of time,
$10^{19}$ times smaller than the nuclear timescale
$t_{proton}=\hbar/m_{proton}c^2=m_{proton}^{-1}$ (translating to the
preferred system of units where $\hbar=c=1$).
The Schwarzschild radius for mass $M$ is $R_S=2M/m_{Planck}^2$;  for
 the
 Sun it  is 2.95 km.} Although the exact value of this ratio
is not critical--- variations of (say) less than a few percent
would not lead to major qualitative changes in the world--- 
neither do structures scale with exact homology, since
other scales of physics are involved in many different
contexts (Carr and Rees 1979).

The maximum number of atoms in any kind of  star is given to order of magnitude by
the large number
\begin{equation}
N_*={M_*\over m_{proton}}\equiv \left({m_{Planck}\over m_{proton}}\right)^3
\approx 2.2\times 10^{57}.
\end{equation}
Many kinds  of equilibria are possible below 
$M_*$ but they are all destabilized 
  above $M_*$  (times a numerical coefficient depending on the structure
and composition of the
star under consideration).
The reason is that above $M_*$ the  particles providing pressure support
against gravity, whatever they are, become relativistic and develop
a soft equation of state which no longer resists collapse;
 far above $M_*$ the only stable compact structures are
black holes.

 A star in hydrostatic equilibrium 
has a size $R/R_S\approx m_{proton}/E$ where the particle energy
$E$ may be thermal   or from degeneracy. Both $R$ and $E$ vary enormously,
for example in main sequence stars thermonuclear burning regulates
the temperature at $E\approx 10^{-6}m_{proton}$, in white dwarfs the
degeneracy energy can be as large as $E_{deg}\approx m_e$ and 
in neutron stars, $E_{deg}\approx 0.1 m_{proton}$.

For example, the Chandrasekhar (1935)
mass, the maximum stable mass of an electron-degeneracy
supported dwarf,
occurs when the electrons become relativistic, at $E\approx m_e$,
\begin{equation}
M_C= 3.1(Z/A)^2M_*
\end{equation}
where $Z$ and $A$ are the average charge and mass of the ions;
typically $Z/A\approx 0.5$ and $M_C=1.4M_\odot$,
where $M_\odot= 1.988\times 10^{33} g\approx
 0.5 M_*$ is the mass of the Sun. 

For main-sequence stars undergoing nuclear burning, the size is fixed by
equating the gravitational binding energy (the typical thermal particle
energy in hydrostatic equilibrium) to the temperature at which 
nuclear burning occurs at a sufficient rate to maintain the outward
energy flux. 
The rate for 
nuclear reactions is determined by  
quantum tunneling through a 
Coulomb barrier by particles on the tail of a thermal
distribution; the rate at temperature $T$ is a
thermal particle rate times
$\exp[-(T_0/T)^{1/3}]$ where 
$T_0=(3/2)^{3}(2\pi Z\alpha )^2 Am_{proton}$. 
Equating this  with a stellar lifetime (see below) yields
\begin{equation}
T\approx (3/2)^3(2\pi)^2\alpha^2m_{proton}[\ln (t_*m_{proton})]^{-3};
\end{equation}
note that the steep dependence of rate on temperature
means that the gravitational binding  energy
per particle, $\propto GM/R$, is almost the
same for all
main-sequence stars, typically 
about $10^{-6}m_{proton}$. The radius of a star is
larger than its Schwarzschild radius $R_S$
by the same factor. Since $M/R$ is fixed, the matter pressure
$\propto M/R^3\propto M^{-2}$  and at large masses (many times $M_*$) is less than the
radiation pressure, leading to instability.

There is a minimum mass for hydrogen-burning stars
 because  electron degeneracy supports a
cold star in equilibrium  with a particle energy $E=m_e(M/M_C)^{4/3}$.
Below  about 0.08 $M_\odot$ the hydrogen never ignites and one has a 
large planet or brown dwarf.
The   maximum radius of a cold planet (above which atoms are gradually crushed
by gravity) occurs where the
gravitational
 binding per atom is about 1 Rydberg, hence $M=M_C
\alpha^{3/2}$--- about the mass of Jupiter.

The same scale governs the formation of stars.  Stars form  from 
interstellar gas clouds in a complex interplay of many scales coupled by 
radiation and magnetic fields,   controlled by transport of radiation
and angular momentum.  Roughly speaking (Rees 1976) the clouds
break up into  small pieces until their radiation is trapped,   when
the total binding energy $GM^2/R$ divided by the gravitational collapse time
$(GM/R^3)^{-1/2}$ is 
equal to the rate of radiation
(say $x$ times the maximum blackbody rate) at $T/m_{proton}\approx GM/R$, giving a
characteristic mass of order $x^{1/2}(T/m_{proton})^{1/4}M_*$,
controlled by the same large number.

Similarly we can estimate lifetimes of stars. 
Massive stars as well as many quasars radiate close to the Eddington
luminosity per mass $L_E/M=3G m_{proton}/2r_e^2=1.25\times
10^{38}(M/M_\odot)$ erg/sec  (at which momentum
transfer by electrons scattering outward  radiation flux balances gravity
on protons),
yielding a  
minimum stellar   lifetime (that is, 
lower-mass stars radiate less and last
longer than this).
The resulting characteristic ``Salpeter time''  is
\begin{equation} \label{eqn:salpeter}
t_*\approx {\epsilon c\sigma_T\over 4\pi G m_{proton}}
=\left[ \epsilon \alpha^2\left({m_{proton}\over m_e}\right)^2\right]
\left({m_{Planck}\over m_{proton}}\right)^3 t_{Planck}\approx
4\times 10^8\epsilon\ {\rm years}
\end{equation}
The energy efficiency $\epsilon\approx 0.007$ for hydrogen-burning stars
and $\approx 0.1 $ for black-hole-powered systems such as quasars.
The minimum timescale of astronomical variability is
the Schwarzschild time at $M_*$,
\begin{equation}
t_{min}
\approx \left({m_{Planck}\over m_{proton}}\right)^2 t_{Planck}
\approx \left({m_{Planck}\over m_{proton}}\right) t_{proton}
\approx \left({m_{Planck}\over m_{proton}}\right)^{-1}t_*
\approx 10^{-5}{\rm sec}.
\end{equation}
The ratio of the two times, $t_*$ and $t_{min}$, which is $\alpha_G^{-1/2}$,
gives the dynamic range of
astrophysical phenomena in  time, the ratio of a stellar evolution time to the
collapse time of a stellar-mass black hole.

A ``neutrino Eddington limit'' can be estimated by replacing 
the Thomson cross section by the cross section for neutrinos
at temperature $T$,
\begin{equation}
L_{E\nu}\approx L_{E}
(m_W/m_e)^2 (m_W/T)^2.
\end{equation}
In a gamma-ray burst fireball or a core collapse supernova,
 a collapsing neutron star releases its
 binding energy $0.1 m_{proton}\approx 100 MeV$ per nucleon, and the neutrino
luminosity 
$L_{E\nu}\approx 10^{54}$ erg/sec liberates the binding energy in a matter
of seconds. This is a rare example of a situation where weak 
interactions and second-generation fermions
 play a controlling role in macroscopic
dynamics, since the energy deposited in the outer layers by neutrinos
is important to the explosion mechanism (as well as nucleosynthesis)
in core-collapse supernovae. 
The neutrino luminosity of a
core-collapse supernova briefly exceeds  the light output of all the stars of universe,
each burst involving $\approx \alpha_G^{1/2}$ of the baryonic mass and lasting
a little more than $\alpha_G^{1/2}$ of the time.

Note that there is a purely relativistic Schwarzschild luminosity 
limit, $c^5/2G=m_{Planck}^2/2=1.81\times 10^{59}$ erg/sec, corresponding
to a mass divided by its Schwarzschild radius. Neither Planck's 
constant nor the proton mass enter here, only gravitational physics.
The luminosity is achieved in a sense by the Big Bang  (dividing radiation
in a Hubble volume by a Hubble time any time 
during the radiation era),   by gravitational radiation during the 
final stages of comparable-mass black hole
mergers,  and continuously by the $PdV$ work done by the negative pressure
of the cosmological constant in a Hubble volume
as the  universe expands.
The brightest individual sources of light, gamma ray bursts, fall four or five
orders of magnitude short of this limit, as does the sum of all   astrophysical
sources of energy (radiation and neutrinos) in the observable universe.

Using cosmological dynamics---
the Friedmann equation  $H^2=8\pi \rho m_{Planck}^2$
relating the expansion rate $H$ and mean density $\rho$---
 one can show that the same number $N_*$ gives
the number of stars within a Hubble volume $H^{-3}$, or that
the optical depth of the universe
to Thomson scattering is of the order of $Ht_*$. The
cosmological connection
between density and time played prominently in Dicke's
rebuttal of Dirac.  Dicke's point  is  that 
the large size and age of the universe--- the reason it is
much bigger than the proton and longer-lived than a nuclear 
collision--- stem from the  large numbers
$M_*/m_{proton}$ and $t_*/t_{proton}$, which in 
turn derive from the large ratio of the Planck mass to the
proton mass. But where does that large ratio come from? Is there
an explanation that might have satisfied Dirac?

\subsection{Running Couplings}

Grand Unified Theories 
point to such an explanation--- a unified model from which one can
derive  the values and ratios of 
the coupling constants.  In these unification schemes,
the three Standard Model
 coupling constants  derive from
 one unified coupling (which is still arbitrary at this level).
The logarithmic running of coupling strength
with energy, derived from renormalization theory,
  leads to the large ratio  
between unification scale and the proton mass. Although
gravity is not included in these theories, the inferred
unification scale ($10^{16}$ GeV) is close   to
the Planck mass; the running couplings thus account for most
of the ``largeness'' of the astrophysical Large Numbers. 

Phenomenological 
coupling constants such as those we have been using
(e.g., $\alpha$) are not really constant but
``run'' or  change with energy scale (Wilczek 1999). 
The vacuum is full of virtual particles which are polarized
by the presence of a charge. An electrical charge 
(or a weak isospin charge) attracts 
like charges, which tend to screen its charge as 
measured from far away. 
At small distances  there is less screening, so the charge
appears bigger, so the effective coupling grows with energy.
On the other hand a strong color charge attracts mostly virtual like-color
charged gluons, so it is antiscreened and the coupling changes
with the opposite sign--- it gets weaker at high energy,
and is said to  display ``asymptotic freedom''.
The freedom 
comes about from the antiscreening by gluons.\footnote{The reason for the
difference is related to the zero point energies being opposite for
fermion and boson modes, which also enters into considerations
about their cancelling contributions to the 
 cosmological constant in supersymmetric vacua.}  

The bookkeeping of how the constants change with the energy scale $M$
of interactions 
is done by renormalization
group calculations. These show that
the running
coupling  constant of U(1), $\alpha_1=g_1^2$, obeys
\begin{equation}
{\partial \alpha_1^{-1}\over\partial \log (M^2)}
=-{1\over 3\pi}\sum Q_i^2
\end{equation}
where the sum is over the charges $Q_i$ of
all fermions of mass less than $M$. The amount of charge screening
by virtual particles increases if the vacuum contains more degrees of 
freedom that can be excited at a given energy.  If all
fermions in the Standard Model are
included (and no more), the total sum on the right side is $14/3$,
yielding a   slope of $-14/9\pi$.

For SU(3), there is again a screening term depending on the number
of color-charged fermions, but there is also an antiscreening
term from the (known number of) gluons,
\begin{equation}
{\partial \alpha_S^{-1}\over\partial \log (M^2)}
={11-(2/3)n_f\over 4\pi}
\end{equation}
where $n_f$ is the number of quark flavors of mass less than $M$.
The factor of 11 from gluons dominates if the number of
quark flavors is not too large, giving asymptotic freedom.
 In the Standard Model,
$n_f=6$, yielding a   slope of $+7/4\pi$.

The running of   couplings depends on
the particle degrees of freedom at each energy scale,
that is, counting virtual particles with rest mass  below
that energy. Thus in reality the slopes change with energy
scale and with the addition of new species, if there are any.

It has been known for over 20 years that the gauge groups
of the Standard Model fit nicely into larger groups of certain Grand Unified
Theories (GUTs),
the   simplest
 ones being SU(5) and SO(10). The coupling constants of $SU(3), SU(2),
U(1)$  all approach each other logarithmically, merging at the GUT scale, about
$10^{16}$ GeV. In recent years     measurements 
of the couplings near $m_Z$ have steadily improved and for some GUTs 
(such as minimal SU(5)) the three couplings no
longer meet at a point; however, the agreement survives impressively
well in supersymmetric models (Langacker and Polonsky 1994),
or in models such as SO(10).
 There is thus some reason to believe that these models work 
 up to the large scale of unification,
which is already close to the Planck mass.

\subsection{Derivation of $m_{Planck}/m_{proton}$}

By the same
token, if one of these GUTs is correct, it will provide
a  derivation of the  
$\alpha_1,\alpha_2,\alpha_3$ coupling constants at any scale from one 
unified constant $\alpha_U$ at the unification scale.
Recall that the mass of the proton is fixed by the scale at
which the SU(3) coupling diverges. Because of the slow
variation of coupling with energy,
 this takes a large range of energy and leads to a
large ratio of proton to unification mass. 

We can run through a toy calculation as follows. Assuming the degrees of
freedom  are constant, the inverse couplings just depend
linearly on the log of the energy scale, so (9)
and (10) can be trivially integrated. Equating them at
the unification scale $M_U$, $\alpha_1(M_U)=\alpha_3(M_U)$, yields  
\begin{equation}
{M_{U}\over {\Lambda}}=\exp\left[{{\alpha_1^{-1}(\Lambda)
-\alpha_3^{-1}(\Lambda)}
\over
{11-(2/3)n_f\over 2\pi}+{2\over 3\pi}\sum Q_i^2}\right]
\end{equation}
Naiively plugging in the standard model numbers (which give 2.1
for the denominator), and the values $\alpha_1\approx (60)^{-1}$ and
$\alpha_3\approx \alpha_S\approx 0.12$
 for
the  coupling constants at the $Z$ scale, yields a mass ratio
of $M_U/M_Z\approx \exp[(60-8)/2.1]=10^{11}$.
This toy estimate is wrong in several details (most notably, not having
included supersymmetry)
 but correctly
illustrates the main point, that there exists an exact calculation
that yields a large ratio of fundamental masses, roughly
\begin{equation}
M_U/m_{proton}\approx
e^{\alpha^{-1}/4}\approx e^{\alpha^{-1}_1(\Lambda_{QCD})/2}
\approx e^{{3\over 2}\alpha_U^{-1} };
\end{equation}
 The  numerical factors   here are just approximate,
but are  exactly computable within the
framework of supersymmetric GUTs  and yield  a unification scale of $M_U\approx 10^{16}$
GeV. In this framework,  this is essentially the explanation of the
``weakness'' of gravity, the smallness of $m_{proton}/m_{Planck}$.
 Since
$m_{Planck}\approx 10^3 M_{U}$   there are  three of
the nineteen orders of magnitude still to be accounted for,
presumably  
by the final unification with gravity. 

Formulae very similar to (12) have appeared for many years
(see, for example, eq. (54) of Carr and Rees 1979).
The rationale has always centered (as it does here) on the logarithmic divergences
of renormalization but in the context of supersymmetry the 
derivation is much crisper--- it comes in the framework of
rigorous derivations in a   well-motivated theory now
being tested (Wilczek 1998).  If this guess about unification is correct, we have
most of the  explanation  of
the large numbers of astrophysics, subject to the value
of one independent, apparently
arbitrary  coupling-constant parameter
($\alpha_U $ or $g_U$),
 a moderately small number  (of the order of 1/25).
The value of $m_{proton}/m_U$ depends exponentially on 
$\alpha_U$ (and hence also on $\alpha$).  
Changes of a few percent
 in the couplings lead to order-of-magnitude variations
in the astrophysical Large Numbers, enough to cause
qualitative change in the behavior of
the astrophysical world.
 The fine structure
constant thereby becomes a  candidate
for selective tuning connected to obtaining 
a suitable strength for gravitation!

\section{ Tuning Light Fermion Masses}

\subsection{Nucleons and Nuclei} 

Like the electronic structure of atoms,
  the basic structure
of neutrons and protons   depend hardly at all on any of the parameters.
 Ignoring for now
the small effect of electric charge and
 quark mass,  proton and neutron structure are the same,
with labels related by isospin symmetry. 
Their internal
structure and mass   are entirely  determined by  strong QCD SU(3) gauge
fields (gluons) interacting with each other and
with the quarks. There are  no adjustable
parameters in the structure, not even a coupling constant, except for the setting
of the 
energy scale. 
\footnote{Ironically,
the nucleon rest mass (which of course includes most of the  rest mass of ordinary
matter) is 99\% dominated  by the kinetic 
energy of the constituents, including roughly
equal contributions from  very light  quarks and massless gluons.} 
Although these nucleon field
configurations are not really ``solved'', the equations which
govern them are known exactly and their structure  
 can be approximately solved
in lattice models of QCD which  correctly estimate
for example the mass ratios of the proton and other hadrons. Basically, the
mass of the proton $m_{proton}= 0.938 GeV$
is some calculable  dimensionless number (about 5) times the 
energy scale
$\Lambda_{QCD}$ fixed by the strong interaction coupling constant.
The structure and mass of hadrons is as mathematically rigid as a Platonic
solid.
 Even so, because $n$ and $p$ are so similar,
 the stability of the proton is very sensitive to the electromagnetic
effects and
to the much  smaller, and seemingly unrelated, up and down quark masses,
which break the symmetry.

Strong interactions   not only create isolated
hadronic structures, but also bind them together into nuclei.
Although the individual hadrons are to first approximation
pure SU(3) solitons, nuclear structure is  also  directly influenced by
  quark masses,
especially through their effect on the range of the nuclear potential. 
The strong interactions of   hadrons can
be thought of as being mediated by  
pions, which have relatively low mass ($m({\pi_0})=$135
MeV) 
and therefore a range which reaches significantly farther than
the hadronic radius.
The light quark masses determine the pion mass via breaking of
chiral symmetry,
  $m_\pi \approx \sqrt{m_{proton}(m_u+m_d)}$, and therefore 
the details of nuclear energy levels are sensitive to 
$u$ and $d$ masses.

The dependence of nuclear structures on quark masses and
electromagnetic forces is hard to compute exactly but we
can sketch the rough scalings.
The nuclear binding energy  in a nucleus with $N$ nucleons
 is about
$E_{nuc}\approx  \epsilon N m_{proton} $ where the specific binding
energy per mass
is about $\epsilon\approx (m_\pi/m_{proton})^2
\approx (m_u+m_d)/m_{proton} \approx 10^{-2} $ and hence
the typical separation is $\epsilon^{-1/2}m_{proton}^{-1}$. 
The nuclear size therefore is typically
 $R\approx N^{1/3} \epsilon^{-1/2}m_{proton}^{-1}$.
 Larger
nuclei  develop increasing electromagnetic repulsion, scaling like 
$E_{em}\approx \alpha  N^2/R $.  They become unstable   above a
maximum charge  where the nuclear and electrostatic 
 energies match,
\begin{equation}
N_{max}\approx
(\epsilon^{1/2}/\alpha)^{3/2}\approx 10^{1.5}.
\end{equation}
The basic reason for the number
of stable nuclei is that the electromagnetic coupling is weak,
but not extremely weak, compared to the strong interactions.

\subsection{Quark masses and the stability of the 
proton and deuteron}

It has long been noted that the stability of the proton   depends on
 the up and down quark masses, requiring
$m_d-m_u \ge E_{em} \approx\alpha^{3/2}m_{proton}$ to overcome the 
extra
electromagnetic mass-energy $E_{em}$ of
 a proton relative to a neutron.
Detailed considerations suggest that $m_d-m_u$ is quite
 finely tuned, in the
sense that if it were  changed by more than a fraction of its value 
either way, nuclear astrophysics as we know it would 
radically change.

Quarks being always confined never appear ``on-shell'' so their masses
are tricky to measure precisely.
 A recent review by Fusaoka and Koide (1998)
gives $m_u=4.88\pm 0.57$ MeV, $m_d= 9.81\pm 0.65$ MeV,
larger than the 0.511 MeV of the electron but  negligible
compared to the 938.272 MeV mass of the proton,
939.566 MeV of the neutron, or 1875.613 MeV of the deuteron.
On the other hand small changes in    $m_d-m_u$ can have
 surprisingly profound effects
on the world through their effect on the relative
masses of the  proton, neutron and deuteron.
 If $m_n<m_p$ the proton is unstable and 
there are no atoms, no chemistry.
It is thus important that $m_n>m_p$, but not by too much since the 
neutron becomes too unstable.
The neutron $\beta -$ decay rate   is as small as it is only because
of the small $n,p$ mass difference: it is 
closely controlled by the phase space suppression.
 With a small increase in the mass difference the neutron
decays much faster and the deuteron becomes unstable,
also leading to radical changes in the world.

Consider for example  the $pp$ reaction,
\begin{equation}
p+p\rightarrow D+ e^++\nu_e,
\end{equation}
which begins the conversion of hydrogen to helium in the Sun.
The endpoint of this reaction is only 420 keV, meaning that if
the deuteron were 420 keV heavier (relative to the other 
reactants)  the reaction would not even be
exothermic
and would tend  to run in the other direction.

Although the  quark masses are  uncertain, we can estimate
 the effect a change in their difference  would have.
To the extent that the neutron and proton  structures
preserve isospin symmetry, the calculation is simple 
since   their masses
 just change additively in response to a change in
the   quark masses.
For the deuteron the story is a little more involved 
because of the effect on the nuclear
potential. 

Consider a transformation to a different world with different values
of the quark and electron masses,
\begin{equation}
%\begin{array}{llll}
m_d\rightarrow m_d'\equiv m_d+\delta m_d,\ 
m_u\rightarrow m_u'\equiv m_u+\delta m_u,\ 
m_e\rightarrow m_e'\equiv  m_e+\delta m_e.
%\end{array}
\end{equation}
We then have
\begin{equation}
%\begin{array}{llll}
m_p'=m_p+2\delta m_u+\delta m_d,\ 
m_n'=m_n+2\delta m_d+\delta m_u,\ 
(m_n-m_p)'= (m_n-m_d)+ \delta m_{d-u}
%\end{array}
\end{equation}
We have defined a key parameter,
the  amount of change  in the mass difference,
$\delta m_{d-u}\equiv \delta m_d-\delta m_u$.

Now consider the effect of this transformation 
 on the reactions
\begin{equation}
n\leftrightarrow p+ e +\bar\nu_e,
\end{equation}
The heat balance of these reactions in our world  is
\begin{equation}
m_n-m_p-m_e-m_{\bar\nu_e}=0.782MeV
\end{equation}
In the transformed world, 
a  hydrogen atom (HI) is unstable (through the proton capturing
the electron and converting into a stable neutron) if
\begin{equation}
\delta m_{d-u}< \delta m_e-0.782\ {\rm MeV}.
\end{equation}
In atoms, or in plasmas where electrons are readily available,
the neutron becomes  the energetically favored state. As 
$\delta m_{d-u}$ drops,  
Big Bang nucleosynthesis first increases the 
helium abundance to near 1, then   makes most of  the baryons
into  neutrons.
There would
be no hydrogen atoms except a small residue of deuterium. 
Synthesis of heavy elements could still continue (although
as shown below, with  the nuclei   somewhat altered). Indeed
there is no Coulomb barrier to keep  the neutrons apart and hardly any
electrons to provide opacity,  so the
familiar equilibrium state of main-sequence stars would disappear.
The effects get even  more radical as $\delta m_{d-u}$ decreases
even more;  rapid,
spontaneous decay of a free proton to a neutron
happens if
\begin{equation}
\delta m_{d-u}< -\delta m_e-2 m_e-0.782\ {\rm MeV}
=-\delta m_e- 1.804\ {\rm MeV}.
\end{equation}
 
For positive  $\delta m_{d-u}$, we have the opposite
problem; neutrons and deuterons are destabilized.
First, we restrict ourselves to constant $\delta m_{d+u}
\equiv \delta m_d+\delta m_u= 0$,
so changes in nuclear potential can be neglected. 
Then we consider just the effect of the change in deuteron mass,
\begin{equation}
m_D'=m_D-\delta m_{d-u} 
\end{equation}
on the   $pp$ reactions
$p+p\leftrightarrow D+ e^++\nu_e$.
In our world the heat balance is
\begin{equation}
2m_p-m_e-m_D-m_{\nu_e}=0.420MeV.
\end{equation}
The $pp\rightarrow D$ direction  stops being energetically favored if
\begin{equation}
\delta m_{d-u}>-\delta m_e +0.42\ {\rm MeV}.
\end{equation}
In the Big Bang plasma, the abundance of deuterons in this world is 
highly suppressed, so  there is no stepping-stone to
the  production of helium and heavier
nuclei, so the universe initially is made of essentially
pure protons.\footnote{The reactions are of course also
affected by couplings which enter into reaction rates. The 
balance between the expansion rate  and weak interaction rates   
controls nucleosynthesis both in supernovae and in the Big Bang.
For example, Carr and Rees (1979) argue that
 avoiding a universe of nearly pure helium  requires the weak
freeze-out to occur at or below the temperature equal to the
$n,p$ mass difference,
requiring
$
(m_n-m_p)^{3}>m_{Planck}^{-1} \alpha^{-2}  m_{proton}^{-2} m_W^{4}.$}  Furthermore, since the $pp$ chain is broken,
cosmic chemical history would be radically altered: For example, 
there is no two-body reaction for nucleosynthesis 
in stars to get started so main-sequence stars
would all have to use catalytic cycles such as the CNO process
(where the heavy catalysts would have to be generated in
an early generation under degenerate conditions).

As long as stable states of heavier nuclei exist, 
some of them would likely be produced occasionally in degenerate
deflagrations (akin to Type Ia supernovae).
As $\delta m_{d-u}$ increases, the valley of $\beta$-stability
 moves to favor fewer
neutrons;  
a free deuteron spontaneously fissions into two protons if
\begin{equation}
\delta m_{d-u}>\delta m_e +0.42\ {\rm MeV}+ 2 m_e
=\delta m_e + 1.442 \ {\rm MeV}.
\end{equation}
Above  some threshold, stable states of heavier nuclei 
disappear altogether and there is no nuclear physics at all.

Thresholds for these effects are shown in figure 1.
Note that   $m_d-m_u$ is bounded within a
small interval--- if it departs from this range  
  one way or another 
 a major change in nuclear astrophysics results.
The total width of the interval, of the order of an MeV,
depending on how drastic the changes are, should be compared
with the values $m_u\approx $5 MeV  and $m_d\approx$10 MeV,
or the mass of the proton, 1 GeV.

We should consider these constraints with the kind of additional 
joint constraints that unification symmetry is likely to impose
on the fermion masses. For example, suppose that  some symmetry
 fixes the ratio $m_d/m_e$ (e.g., Fukugita et al. 1999),
thereby fixing $\delta m_d/\delta m_e$, and we require
that $m_u>0$. 
The resulting constraint is illustrated in figure 1.

\subsection{Quark masses and the range of nuclear forces: diproton
 stability}

We have explored two of the three dimensions in 
$\delta m_u,\delta m_d,\delta m_e$ space: $\delta m_d-\delta m_u$
and $m_e$.  In addition
 there is a  third dimension
to explore, $\delta m_d + \delta m_u$. This quantity affects the 
pion mass and therefore the range of the nuclear interactions;
this does not affect the $np$ stability arguments
but does affect the $D$ stability.

The dependence on this
third dimension of fermion mass variation can be estimated
 through the  effect of changes in
 nucleon potential through the pion mass, 
$m_\pi^2\propto (m_u+m_d)\Lambda_{QCD}$. In this framework
Agrawal et al. (1998) investigated the effect of 
varying the Higgs expectation value $v$, which changes all the fermion
masses   in proportion.  Using a  simple model of 
the deuteron potential (range 2 fm, depth 35MeV)
they found  no bound
states anymore if the range is reduced by 20\%, or the quark
mass sum is  increased by 40\%. This   corresponds to a change
$v/v_0=1.4$ or $\delta m_i=0.4 m_i$, or approximately
$\delta m_d+\delta m_u\approx 0.4 (m_d+m_u)\approx 7$ MeV.
 (See also the earlier discussion of light nuclei
stability by Pochet et al.  1991).
On the side of decreasing quark masses
or increasing range (i.e. $\delta m_d + \delta m_u
<0$), the effects are opposite; at about
$\delta m_d +\delta m_u\approx -0.25 (m_d+m_u)\approx -4$MeV,
the diproton $^2$He or the dineutron become bound (Dyson 1971). (Which one
is stable depends on the mass difference  $\delta m_{d-u}$.)
 However, a tighter constraint in this dimension is likely to arise from
the behavior of heavier nuclei.

\subsection{ Tuning levels of heavier nuclei}

The most
celebrated nuclear tunings, first noticed by Salpeter  and Hoyle,
  involve the resonant
levels of  carbon and oxygen nuclei. The excited resonance
 level of $^{12}C^*$ at
7.65 MeV lies just 0.3MeV
 above the 7.3667 MeV energy of $^{8}Be+^4He$, allowing
rapid enough reactions for carbon to form before the
unstable $^8Be$ decays. On the other hand the level of 
$^{16}O$ at 7.1187 MeV lies just below that of $^{12}C+^4He$
at 7.1616 MeV; if it were higher by
just 0.043 MeV, reactions to oxygen would  quickly destroy the carbon. 
The way these interlocking levels depend on $m_d,m_u,m_e$ is
too hard to compute from first principles in detail,  but 
Jeltema and Sher (1999) have recently estimated the effect on the
nuclear
potential  
of  adjusting the
Higgs parameter $v$,  tracing its effect on the first reaction
above  through
the work of Oberhummer et al. (1994) and Livio et al. (1989). 
In this way they estimate a lower bound $v/v_0>0.9$. 
Oberhummer et al. (1999) have recently computed the
dependence of these levels in a simple cluster model
for the nuclei, and conclude that  the
strength of the nuclear force needs to be tuned to 
 the 1\% level.  This can be interpreted to mean that the  products would be radically
altered if $\delta m_u+\delta m_d$ changed by even a few percent of $m_u+m_d$,
on the order of  0.05 MeV.

\section{Fixed and Adjustable Parameters in the Final Theory}

The  structural properties of the world are not sensitive to small
local perturbations of many  parameters about their actual
values. However,    nuclear physics
would change drastically with even small changes in
 $  m_u$ and  $m_d$ at the level of a few percent.
Grand unification leaves these as independent parameters
  without relations fixed by symmetries, so we
may conjecture that they remain so in  
more inclusive unified theories. This leaves just about
enough freedom for a multiverse to find a world which has stable 
protons, produces carbon and oxygen, and still endows these
atoms with a rich interactive chemistry.

The paradigm of a fixed, calculable, dimensionless quantity in physics is the
 anomalous magnetic
moment of the electron  (Hughes and Kinoshita 1999).
 In a display of spectacular experimental and 
theoretical  technique,
it is measured to be (Van Dyck, Schwinberg and Dehmelt 1987)
$a\equiv(g-2)/2=1,159,652,188.4 (4.3)\times 10^{-12}$,
a precision of 4 parts per billion; it is calculated to 
even better accuracy except for the uncertainty in the fine structure
constant, which limits accuracy of the agreement to about 30 ppb. 
This agreement cannot be an accident--- the   precision tells
us that we really understand the origin of this dimensionless number.
The precision is exceptional because the 
dimensionless numbers can be measured so accurately and  the
theory is clean enough to calculate so accurately.
It is hard to measure precisely because nothing in particular
  depends critically on
what the exact final digits  in the expansion are.
 We expect this to be so in such
a case of a  mathematically computable number. It would
be disturbing if a   different number in the ninth decimal place would make
a big difference to (say) element production, because it would indicate 
 a conspiracy at a level where we have no mechanism to explain it.
On the other hand a fine tuning in an adjustable parameter is easy
to live with because we have a physical way to arrange that. So,
the attitude adopted here is that  maybe we can find
the adjustable parameters by looking for the
places where fine  tuning is needed. The clue is in the derivative
$\Delta$World/$\Delta$parameter, how much the phenomena change as 
a result of a parameter change; 
 we should look for the fundamental  flexibilities
in the fundamental theory where this derivative is large.  

Grand Unification   permits about enough  freedom
 in Standard Model parameters to account for
the apparent fine tunings by selection from an ensemble of 
possibilities. This
 is a useful lesson to bear in mind as unification theory forges
 ahead seeking to fix new predictions---
contrary to the aspirations of many in the unification
community, we should not expect   to find  more relationships
among Standard Model parameters to be    fixed by symmetry
in the final theory  than are  
fixed by the ideas we have in place already, at least not
among the light fermion masses.

 These considerations may   help to guide us to the connections of 
superstrings to the low energy world. For example,  Kane et al. (2000) have pointed out
that the ideal superstring  theory indeed predicts absolutely everything,
including the light lepton mass ratios, seemingly allowing
no room for tuning. However, even here there is
the possibility that the exact  predictions do  not specify 
a unique universe at low energy but correspond  
to  many  discrete options--- many minima in a vast superpotential. If the 
minima are numerous enough a close-to-optimal set of parameters can still
be found. The fundamental theory might  predict  the properties of all the minima
but the main choice may still be made by selection. String-motivated ideas for
explaining  the mass hierarchy outside of the context of standard GUTs
 (e.g. theories with extra dimensions--- Arkani-Hamed et al. 1998, Dienes et al. 1998,
1999, Randall and Sundrum 1999)
may offer similar options for optimizing
 the Yukawa couplings.

Anthropic arguments are often said to lack
predictive power. However, within a theoretical
framework  specific predictions do emerge
from the guesses made from anthropic clues, which could falsify 
a particular conjecture: for example, the conjecture that
the deuteron and proton stability arise from selection of
light quark masses from a continuous spectrum of possible values
predicts 
that in fundamental theory, it will not be possible to mathematically
derive from first principles the value of $(m_d-m_u)/m_{proton}$.
 At the very 
least this should be regarded as a challenge to a community which has
so far been very successful in discovering ways to reduce the number
of free parameters  in various unification schemes. One is reminded
of Darwin's theory, which is a powerful explanatory tool 
 even though some question its  predictive  power.
Anthropic arguments are  vulnerable in the same way  to ``just-so'' 
storytelling but may nevertheless form an important part of
cosmological theory.

\section*{ACKNOWLEDGMENTS}

I am grateful for conversations with  D. Brownlee,
S. Ellis,  G. Gonzalez, W. Haxton, G. Kane, M. Perry,
 J. Preskill, M. Rees, S. Sharpe, M. Tegmark,  P. Ward, and L. Yaffe, and
  am particularly grateful for detailed comments by M.
Fukugita.  This
work was supported by NASA and NSF  at the University of Washington,
by the Isaac Newton Institute for Mathematical Sciences, University
of Cambridge, and  
by a Humboldt Research Award at the Max Planck Institute f\"ur Astrophysik,
Garching.

\begin{figure}[t]
\caption{
Effects of changes in the light  quark mass difference and
electron mass on the stability of the proton and deuteron.
Our world sits at the origin; outside the bold lines 
 nuclear 
astrophysics changes qualitatively in four ways described in the text.
The physical effects are: destabilization of an isolated deuteron;
destabilization of a proton in the presence of an electron;
$pp$ reaction goes the wrong way; destabilization of an 
isolated proton. 
Thresholds  are shown for the four
effects--- solid lines from equations 19  and
21, dashed lines from  23 and  24, 
the latter assuming $\delta m_d+\delta m_u=0$.
Dotted lines
 show a constraint (appropriate in an SO(10) GUT) imposed by positive up-quark
mass for fixed $\delta m_e/\delta m_d$,  and we plot only the region of
 positive electron mass. 
The change in the sum  $\delta m_{d+u}$ (the combination
not shown here) is similarly constrained within
less than 0.05 MeV of its actual value  so as not to drastically alter
carbon-producing reactions.}
\end{figure}


\begin{references}
\bibitem{}
Acu\~na et al. 1999, Science 284, 790
\bibitem{}
Agrawal, B., Barr, J. F., Donoghue, J.F. and Seckel, D. 1998,
Phys Rev Lett 80, 1822
\bibitem{}
Agrawal, B., Barr, J. F., Donoghue, J.F. and Seckel, D. 1998,
Phys Rev Lett D57,5480
\bibitem{}
Arkani-Hamed, N., Dimopoulos, S., and Dvali, G. Phys. Lett. B429, 263
%,hep-ph/9803315
\bibitem{}
Berezhiani, Z. 1996, in Proceedings of ICTP summer school, hep-ph/9602325
\bibitem{}
Barrow, J. and Tipler, F. 1986, {\it The Anthropic Cosmological Principle}
(Clarendon Press, Oxford)
\bibitem{}
de Bernardis, P. et al. 2000, Nature 404, 955
\bibitem{}
Cahn, R. N. 1996, Rev. Mod. Phys 68, 951
\bibitem{}
Carr, B. J. and Rees, M. J., 1979. Nature 278,611
\bibitem{}
Carter, B. 1974. in {\it Confrontation of Cosmological Theories
with Observational Data}, IAU symp. 63, ed. M. Longair (Reidel: Dordrecht)
\bibitem{}
Carter, B. 1983. Phil. Trans. Roy. Soc. A 310, 347
\bibitem{}
Cen, R. and Ostriker, J. P. 1999, Astrophys. J.  519, 1
\bibitem{}
Chandrasekhar, S. 1935, Mon. Not. R. astr. Soc. 95, 207
\bibitem{}
Connerney, J. E. P. et al. 1999, Science 284, 794
\bibitem{}
Dicke, R. H. 1961, Nature 192, 440
\bibitem{}
Dienes, K. R., Dudas, E. and Gerghetta, T. 1998
Phys.Lett. B436,  55
\bibitem{}
Dienes, K. R., Dudas, E. and Gerghetta, T. 1999
Nucl.Phys. B537, 47
\bibitem{}
Dirac, P. A. M. 1938, Proc. Roy. Soc. 165A, 199
\bibitem{}
Doolittle, W. F. 1999,  Science 284, 2124
\bibitem{}
Doyle, L. R., ed., 1996, {\it Circumstellar Habitable Zones},
(Menlo Park: Travis House)
\bibitem{}
Dyson, F. J. 1971, Scientific American, 225 (September), 51
\bibitem{}
Efstathiou, G. 1995, Mon. Not. Roy Astro. Soc. 274, L73
%1996  in
%Cosmological Constant and the Evolution of the Universe. Edited by
%K. Sato, T. Suginohara, and T. Sugiyama. Tokyo, Japan:
%Universal Academy Press, 1996 ., p.225
\bibitem{}
Evans, D. A., Beukes,  N. J. and Kirschvink, J. L. 1997, Nature 386, 262
\bibitem{}
Fukugita, M. 1999, in {\it Structure Formation in the Universe},
Proc. NATO ASI, Cambridge, astro-ph/0005069
\bibitem{}
Fukugita, M., Hogan, C. J. and Peebles, P. J. E. 1996, Nature 381, 489
\bibitem{}
Fukugita, M., Tanimoto, M. and Yanagida, T. 1999, Phys Rev D 59, 113016
\bibitem{}
Fusaoka, H. and Koide, Y 1998, Phys. Rev. D 57, 3986
\bibitem{}
Gaillard, M. K., Grannis, P. D. and Sciulli, F. J. 1999,
Rev. Mod. Phys. 71, S96
\bibitem{}
Garriga, J., Livio, M.  and Vilenkin, A. 2000, Phys.Rev. D61  023503
\bibitem{}
Gonzalez, G., 1999, preprint, University of Washington
\bibitem{}
Hanany, S. et al. 2000, ApJ submitted, astro-ph/0005123
\bibitem{}
Heisler, J. and Tremaine, S. 1986, Icarus 65, 13
\bibitem{}
Henderson, L. J. 1913, {\it The Fitness of the Environment;
 an inquiry into the biological significance of the properties of matter},
  Boston, Beacon Press (1958)
\bibitem{}
Hoffman, P.
 Kaufman, A., Halverson, G. and Schrag, D. 1998, Science 281, 1342
\bibitem{}
Hoyle, F., Dunbar, D. N. F., Wenzel, W. A. and Whaling, W. 1953, Phys Rev 92, 1095
\bibitem{}
Hughes, V. W. and Kinoshita, T. 1999, Rev. Mod. Phys. 71, S133
\bibitem{}
Jeltema, T. and Sher, M. 1999, hep-ph/9905494
\bibitem{}
Kane, G.   Perry, M., and Zytkow, A., preprint  astr0-ph/0001197
\bibitem{}
Kasting, J. F., Whitmire, D. P. and Reynolds, R. T. 1993, Icarus 101,
108
\bibitem{}
Knoll, A. H. 1999, Science 285, 1025
\bibitem{}
Kolb, E. W. 1997, in {\it Astrofundamental Physics}, ed. N. Sanchez
and A. Zichichi, (World Scientific), p. 162, astro-ph/9612138
\bibitem{}
Langacker, P. and Polonsky, N. 1994, Phys Rev D 49, 1454
\bibitem{}
Leslie, J. 1989, {\it Universes}, New York: Routledge
\bibitem{}
Lilly, S. et al. 1998, Astrophys. J.  500, 75
\bibitem{}
Linde, A. 1998, Phys. Rev D. 58, 083514, gr-qc/9802038
\bibitem{}
Livio, M., Hollowell, D., Weiss, A. and Truran, J. W. 1989, 340, 281
\bibitem{}
Madau, P. 1999, Physica Scripta, in press, astro-ph/9902228
\bibitem{}
Martel, H., Shapiro, P. and Weinberg, S. 1998,
Astrophys. J.  497, 29% astro-ph/9701099
\bibitem{}
Matese,  J. and Whitmire, D. 1996, Astrophys. J. 472, L41
\bibitem{}
Miller, A. D. et al. 1999, astro-ph/9906421
\bibitem{}
Oberhummer, H. et al. 1994, Z. Phys. A 349, 241
\bibitem{}
Oberhummer, H., Csoto, A. and Schlattl, H. 1999, astro-ph/9908247
\bibitem{}
Perlmutter, S. et al. 1999, Astrophys. J.  517, 565
\bibitem{}
Pochet, T., Pearson, J. M., Beaudet, G. and Reeves, H. 1991,
Astron. Astrophys. 243, 1
\bibitem{}
Randall, L. and   Sundrum, R. 1999 %``A Large Mass Hierarchy from a Small Extra
Dimension,'' Phys.Rev.Lett. 83,   3370 %, hep-ph/9905221. 
\bibitem{}
Rees, M. J. 1976, Mon. Not R. astr. Soc. 176, 483
\bibitem{}
Rees, M. J.,1997, {\it Before the Beginning} (Reading: Perseus)
\bibitem{}
Riess, A. G. et al. 1998, Astron. J. 116, 1009
\bibitem{}
Rozenthal, I. L. 1980, Sov. Phys. Usp. 23(6), 296
\bibitem{}
Salpeter, E. E. 1964, Astrophys. J. 140, 796
\bibitem{}
Schwartzmann, D. W. and Volk, T. 1989, Nature 340, 457
\bibitem{}
Tegmark, M. 1998, Annals Phys. 270, 1 
\bibitem{}
Tegmark, M. and Rees, M. J. 1998, ApJ 499, 526, astro-ph/9709058
\bibitem{}
Turok, N. and Hawking, S. 1998, Phys. Lett. B432, 271, hep-ph/9803156
\bibitem{}
Vilenkin, A. 1995 Phys.Rev. D52, 3365
\bibitem{}
Vilenkin, A. 1998a, Phys. Rev. D 57, 7069, hep-ph/9803084
\bibitem{}
  Vilenkin, A.   1998b, Phys. Rev. Lett. 81, 5501
gr-qc/9806185
\bibitem{}
Ward, P. and Brownlee, D. 1999, {\it Rare Earth}, (New York: Springer)
\bibitem{}
Weinberg, S. 1987, Phys Rev Lett 59, 2607
\bibitem{}
Weinberg, S. 1989, Rev Mod Phys 61, 1
\bibitem{}
Weinberg, S. 1997, in {\it Critical Dialogues in Cosmology},
ed. N. Turok (World Scientific), p. 195
\bibitem{}
Weinberg, S. 2000, preprint, astro-ph/0005265
\bibitem{}
Wilczek, F. 1998, hep-ph/9809509
\bibitem{}
Wilczek, F. 1999, Rev. Mod. Phys. 71, S85
bibitem{}
Woese, C. R., Kandler, O. and Wheelis, M. L. 1990, Proc. Nat. Acad. Sci.
USA 87, 4576
\bibitem{}
Zlatev, I., Wang, L. and Steinhardt, P. J. 1999,
Phys. Rev. Lett. 82, 896 
\end{references}
\end{document}